\documentclass[twocolumn,aps,preprintnumbers,prl,showpacs,
               amsfonts,amsmath,amssymb,superscriptaddress,floatfix]{revtex4}
\usepackage[dvips]{graphicx}

\begin{document}
\title{Invariance of the bit error rate \\
       in the ancilla-assisted homodyne detection}
\date{\today}
\author{Yuhsuke Yoshida}
\author{Masahiro Takeoka}
\author{Masahide Sasaki}
\affiliation{%
     Quantum Information Technology Group, \\
     National Institute of Information and Communications Technology
     (NICT), \\
     4-2-1 Nukui-kitamachi, Koganei, Tokyo 184-8795, Japan}
\begin{abstract}
We investigated the minimum achievable bit error rate of discriminating 
binary coherent states by homodyne measurements with the help of
arbitrary ancilla states coupled via beam splitters. 
We took into account arbitrary pure ancillary states 
those were updated in real-time by those partial measurements 
and classical feedforward operations where the partial measurements 
were made by homodyne detections with a common fixed phase. 
It was shown that the minimum bit error rate of the system was invariant
under these operations.
We also discussed how generalize the homodyne detection beyond
the scheme which made the bit error rate invariant. 

\end{abstract}
\maketitle

\section{Introduction}

Coherent states are fundamental information carriers
in optical and quantum communications.
They transmit in pure state intact even through a lossy channel.
A primitive modulation scheme is the binary phase shift keying (BPSK)
with $|\pm\alpha\rangle$.
Because these two signals are not orthogonal, i.e.
$\langle\alpha|-\alpha\rangle\ne0$,
it has been a central issue in quantum communications
how to realize a quantum receiver that can discriminate
them as small error as possible.

Toward approaching the minimum bound
for the average error probability (the bit error rate, BER),
the so-called Helstrom bound 
\cite{Hels}, 
optimal 
\cite{Doli,UsHi,SUH,SaHi,TSLL} 
and 
near-optimal 
\cite{Kenn,TaSa,WTCS} 
receivers have been studied theoretically. 
Some of them have been put into experimental demonstrations 
\cite{CMG,WTCS}.

These experiments were based on photon counters 
with additional feedback \cite{Doli,TSLL,CMG} 
or coherent displacement operations \cite{Kenn,TaSa,WTCS}.  
However, due to severe limitations on their detection efficiency, 
the results demonstrated so far are still far from 
the theoretical limit without compensating the detection efficiencies. 
Although there is a rapid progress in the development of 
highly efficient photon counters \cite{LMN,Fukuda09}, 
these are still very advanced technologies and 
not always available in quantum optics labs.

Another important attempt is to use homodyne detectors. 
Homodyne detection is already an well matured technique,
and is used to discriminate the BPSK signals 
in conventional optical communications. 
Some quantum receiver schemes have been proposed in which 
a homodyne detector is combined 
with additional nonlinear processes, 
such as Kerr medium, or higher nonlinear media 
\cite{UsHi,SUH,SaHi}. 
It is, however, still a formidable task to implement 
highly nonlinear unitary processes with low losses.

Instead of direct nonlinear unitary processes,
measurement induced nonlinear process is another attractive possibility 
to combine with homodyne detectors. 
Such a process is made of non-classical ancillary states,
linear optical circuit, homodyne detector, and classical feedforward system.
It was already known that Gaussian operations and classical 
feedforward alone cannot overcome the homodyne BER limit 
\cite{TaSa}. 
Therefore non-Gaussian elements, or non-Gaussian ancillary inputs,
are required essentially. 
Such states are currently available in the laboratory 
\cite{OTLG, NMHMP, WTFS, OJTG, TWSTHFS}. 
It would then be an interesting question 
whether the BER can be reduced below the homodyne BER limit by
introducing non-Gaussian ancillary inputs to 
the homodyne-feedforward linear optical circuit.

In this paper, 
we investigate the achievable BER for the BPSK signals
when one is available to use arbitrary pure ancillary state, 
beam splitter, homodyne detector,
and updating the ancillary inputs based on the homodyne results. 
We show that under any updating of the beam splitter parameters 
and the ancillary inputs, 
the BER is invariant, 
being exactly the same as the homodyne limit. 
Our result suggests how to set up the system with homodyne detection
in order to overcome the basic homodyne limit.

\section{multi-ancilla recombinations}

\begin{figure}
\begin{center}
\includegraphics[width=0.9\linewidth]{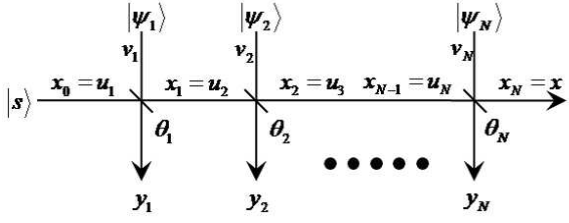}
\caption{\label{fig:updating_scheme}
The received signal $|s\rangle$ is combined 
with the ancillary states $|\psi_1\rangle$, ..., $|\psi_N\rangle$ 
successively 
via the beam splitters with transmittance $\cos\theta_n$.
Both $|\psi_n\rangle$ and $\theta_n$ are updated
based on the homodyne measurement results of ancillae.
The signal is finally discriminated by a homodyne detector. 
The variables $x_{n-1} = u_n$ correspond to the quadrature values 
used for the expansion of the transformed signal state 
in the $n$-th step.
}
\end{center}
\end{figure}

Let us show our system in FIG.~\ref{fig:updating_scheme}.
The received signal $|s\rangle$ 
($s = \pm \alpha$, 
and without loss of generality $\alpha$ is real) is combined 
with the first ancillary state $|\psi_1\rangle$ via a beam splitter 
of a transmittance $\cos\theta_1$, the first homodyne detector, 
outputting a result $y_1$. 
This result is used for updating the second ancilla and 
the beam splitter transmittance. 
Any explicit expression of the updating rule is not needed 
to show the BER invariance.

To describe the updating, we introduce notations. 
Ancillary state $|\psi\rangle$ is expanded by the coherent state as 
\begin{equation}
| \psi \rangle =
    \int \frac {d\alpha}{\pi}
| \alpha \rangle \psi(\alpha),
\label{eq:ancilla integral}
\end{equation}
where $\psi(\alpha)\equiv \langle\alpha|\psi\rangle$, and
$d\alpha \equiv d\alpha_R \wedge d\alpha_I$ for
the decomposition $\alpha = \alpha_R + i\alpha_I$ with
$\alpha_R, \alpha_I \in {\bf R}$. 
Denoting the signal port 
(the horizontal line in Fig.~\ref{fig:updating_scheme}) 
by the subscript ``$0$", 
and the $n$-th ancillary port by the subscript ``$n$", 
the whole $(n+1)$-mode state up to the $n$-th step is recursively defined by 
\begin{eqnarray}
&&
|\Psi_n\rangle_{0,1,\cdots,n} \equiv
    \hat{B}_{0,n}(\theta_n) |\Psi_{n-1}\rangle_{0,1,\cdots,n-1} \otimes
    |\psi_n\rangle_n \nonumber \\
&& \!\!\!\!\!\!\!\!\!\! =
    \int\frac{d^n\alpha}{\pi^n}
    | s'_n      \rangle_0 \otimes
    | \alpha'_1 \rangle_1 \psi_1(\alpha_1) \otimes \cdots \otimes
    | \alpha'_n \rangle_n \psi_n(\alpha_n).
\end{eqnarray}
The action of the beam splitter $\hat{B}_{0,n}(\theta) \equiv
  \exp[\theta(\hat{a}_0^\dagger\hat{a}_n - \hat{a}_n^\dagger\hat{a}_0)]$
between the $0$-th and $n$-th port is given by the matrix representation 
\begin{eqnarray}
\left(\begin{array}{c}
               s'_n \\
          \alpha'_n
      \end{array}\right)
&\equiv&
R(\theta_n)
\left(\begin{array}{c}
          s'_{n-1} \\
          \alpha_n
      \end{array}\right), \\
R(\theta) &\equiv& \left(\begin{array}{c c}
     \cos\theta &  \sin\theta \\
    -\sin\theta &  \cos\theta
    \end{array}\right), \label{eq:R}
\end{eqnarray}
where $s'_0 \equiv s$. 
Here we notice that the amplitude $s'_n$ depends on all the 
$n$ coherent amplitudes 
$\boldsymbol \alpha^{(n)} \equiv (\alpha_1,\cdots,\alpha_n)$.

Now let $y_n$ ($n=1, \cdots, N$) be the $n$-th homodyne result 
and $x$ be the homodyne result at port ``$0$".
The parameter $x$ is a eigenvalue of the quadrature amplitude 
$\hat{x} \equiv (\hat{a} + \hat{a}^\dagger)/\sqrt{2}$, namely 
$\langle x|\hat{x} = x\langle x|$. 
The wave function is then 
\begin{eqnarray} 
&& \Psi_N(x,{\bf y}^{(N)}) \equiv
          {}_0\langle x|\otimes{}_1\langle y_1|
          \otimes \cdots \otimes
          {}_N\langle y_N| \Psi_N \rangle_{0,\cdots,N}
\nonumber \\ &&
     = \int\frac{d^N\alpha}{\pi^N}
       \psi_1 \cdots \psi_N
\times
         \langle x   | s'_N      \rangle
         \langle y_1 | \alpha'_1 \rangle
 \cdots  \langle y_N | \alpha'_N \rangle \label{eq:waveN}
\end{eqnarray}
with 
\begin{equation}
\langle x\mid \alpha\rangle \equiv
      \frac{1}{\sqrt[4]{\pi}}
      \exp\left[   \frac{\alpha^2 - |\alpha|^2}{2}
             - \frac{1}{2}\left(x-\sqrt2\alpha\right)^2 \right].
\label{eq:wave0}
\end{equation}

The probability density of Eq.~(\ref{eq:waveN}) is 
\begin{eqnarray}
&& P(x,{\bf y}^{(N)}|s)
    \equiv
    |\Psi_N(x,{\bf y}^{(N)})|^2 =
\nonumber \\ &&
\int\frac{ d^N\alpha d^N\beta }
         { \pi^\frac{5N+1}{2} }
    \psi^\ast_1(\alpha_1) \psi_1(\beta_1) \cdots
    \psi^\ast_N(\alpha_N) \psi_N(\beta_N) \nonumber \\ &&
\times
    \exp \left[ J - ( x   - s''_N      )^2
      - \sum_{n=1}^N ( y_n - \alpha''_n )^2 \right],
\label{eq: total probability}
\end{eqnarray}
where
\begin{eqnarray}
J &\equiv&
\frac{1}{2}\sum_{n=1}^N \left[ 
    ( \alpha^\ast_n \beta_n - \beta^\ast_n \alpha_n )
  - \left| \alpha_n - \beta_n \right|^2
\right], \\
\left( \begin{array}{c}
       s''_n \\
  \alpha''_n
\end{array} \right)
&\equiv&
    \frac{1}{\sqrt{2}} \left(
    \begin{array}{c}
         s'^\ast_n ({\boldsymbol \alpha^{(n)}})
                 + s'_n ({\boldsymbol  \beta^{(n)}}) \\
       \alpha'^\ast_n ({\boldsymbol \alpha^{(n)}})
            + \alpha'_n ({\boldsymbol  \beta^{(n)}})
    \end{array} \right).
\end{eqnarray}

\section{Bit error rate without feedforward}

We first evaluate the BER 
in the case where the signal is discriminated by 
the $(N+1)$ outcomes of the homodyne detectors 
${\bf y}^{(N)}$ and $x$ 
without any feedforward in Fig.~\ref{fig:updating_scheme}. 
Putting $x_N \equiv x$, 
we transform the output variables $(x_n, y_n)$
into the new ones $(u_n, v_n)$ by applying 
\begin{equation} \label{eq:Rinv}
\left( \begin{array}{c} u_n \\ v_n \end{array} \right)
= R(\theta_n)^{-1}
\left( \begin{array}{c} x_n \\ y_n \end{array} \right). 
\end{equation}
in a descending order $n = N, N-1, \cdots, 2, 1$, 
iteratively putting $x_{n-1} \equiv u_n$. 
The quadratic parts in Eq.~(\ref{eq: total probability}), say $\Omega^{(N)}$,
is invariant under all of the rotational operations (\ref{eq:R}),
and then is equivalent to 
\begin{eqnarray*}
\Omega^{(N)}
&=&
    \left( x_N - s''_N \right)^2 +
         \sum_{n=1}^N \left( y_n - \alpha''_n \right)^2
\\ &=&
\left( x_0 - \sqrt{2}s \right)^2
  + \sum_{n=1}^N \left( v_n
  - \frac{ \alpha^\ast_n + \beta_n }{\sqrt{2}} \right)^2. 
\end{eqnarray*}
Noting that the Jacobian of this transformation is
$\prod_{n=1}^N \det R(\theta_n) = 1$, the conditional probability is now expressed in terms of 
the new variables $x_0, {\bf v}^{(N)}=(v_1, \cdots, v_N)$ as
\begin{eqnarray}
P(x  , {\bf y}^{(N)} |s)    
\rightarrow P(x_0, {\bf v}^{(N)} |s)  
=
P(x_0|s)  P({\bf v}^{(N)}| s),
\end{eqnarray}
where
\begin{eqnarray}\label{eq:factorization}
&& P(x_0|s) \equiv
\frac{1}{\sqrt{\pi}} \exp \left[ - (x_0 - \sqrt{2}s)^2 \right],
\nonumber\\
&& \!\!\!\!\!\!\!\!\!
P({\bf v}^{(N)}| s) \equiv \int\frac{d^N\alpha d^N\beta }
              {\pi^\frac{5N+1}{2}}
              \psi^\ast_1(\alpha_1) \psi_1(\beta_1)
              \cdots \psi^\ast_N(\alpha_N) \psi_N(\beta_N) \nonumber \\
&&
   \times \exp \left[
         J - \sum_{n=1}^N \left( v_n
       - \frac{\alpha^\ast_n + \beta_n}{\sqrt{2}} \right)^2
     \right].
\end{eqnarray}
Thus it can be factorized into two parts, 
and  
the probability of obtaining the value 
$x_0$ is independent on those of ${\bf v}^{(N)}$.

Therefore the BER can be calculated independent of 
the ancillary homodyne outcomes. 
So, by introducing the threshold
\begin{equation}
x_{th} \equiv \frac{1}{4\sqrt{2}\alpha}\ln\left(
              \frac{p(-\alpha)}{p(\alpha)}\right) \label{eq:threshold}
\end{equation}
determined by 
$P(x_{th}|\alpha) p(\alpha) = P(x_{th}|-\alpha)p(-\alpha)$ 
in a priori probability $p(\pm\alpha)$, 
the BER is given by 
\begin{eqnarray}
&& BER_N \equiv 
\int_{-\infty}^\infty d{\bf v}^{(N)} 
\Biggl[
       \int_{-\infty}^{x_{th}} dx_0
       P(x_0, {\bf }v^{(N)}| \alpha) p(\alpha) \nonumber \\
&& \quad + \int_{x_{th}}^\infty    dx_0
       P(x_0, {\bf v}^{(N)}| -\alpha) p(-\alpha)
   \Biggr] \nonumber \\
&& =
\int_{-\infty}^{x_{th}}\!\!\!dx_0 P(x_0|\alpha) p(\alpha)
 +
\int_{x_{th}}^{\infty}\!\!\!dx_0 P(x_0|-\alpha) p(-\alpha) \nonumber \\
&\equiv& BER_0, \label{eq:BERN}
\end{eqnarray}
which is identical with the homodyne limit. 
For the equivalent a priori probabilities $p(\alpha) = p(-\alpha)=1/2$,
we have $BER_0 = {\rm erfc}(\sqrt{2}\alpha)/2$ as is well known.


\section{Bit error rate with feedforward}

Next we consider generic feedforward  
by updating ($\psi_n$, $\theta_n$) based on the homodyne outcomes 
${\bf y}^{(n-1)}$. 
Suppose the parameters ($\psi_n$,$\theta_n$) are functions of
the previously measured values ${\bf y}^{(n-1)}$
of the homodyne detections,
and then are regarded as functions of $v_1$, $\cdots$, $v_{n-1}$ 
through the rotation (\ref{eq:Rinv}).
After changing variables via this rotation (\ref{eq:Rinv}),
the generic functional relations become
\begin{eqnarray}\label{eq:func-rel}
     \psi_n &=&      \psi_n( x_0, {\bf v}^{(n-1)}), \nonumber \\
\theta_n &=& \theta_n( x_0, {\bf v}^{(n-1)}),
\end{eqnarray}
for $n = 2, 3, \cdots, N$.
These are shown by eliminating ${\bf y}^{(n-1)}$
from $(\psi_n({\bf y}^{(n-1)}), \theta_n({\bf y}^{(n-1)}))$
by iteratively applying
\begin{eqnarray*}
y_m &=& v_m     \cos\theta_m + x_{m-1} \sin\theta_m, \\
x_m &=& x_{m-1} \cos\theta_m - v_m     \sin\theta_m,
\end{eqnarray*}
for $m=n-1, n-2, \cdots, 2, 1$.

The parameters ($\psi_1$,$\theta_1$) are given prior to any ancillary states.
The $BER_N$ will be calculated with the help of the relation
(\ref{eq:func-rel}) in the feedforward control case.
Note that although the probability density of $y_n$
depends on the value of signal $s$,
there is no functional relation with the parameters
($\psi_n$,$\theta_n$) to the value of signal $s$.
The signal $s$ affects the value of $y_n$
only through changing its probability distribution.
The functional relations (\ref{eq:func-rel}) show that ancillary states
do not depend directly on the input signal $s$,
but depend only through probability densities of $y_n$'s.
This fact helps us to prove the invariance of the BER for
the updating operation considered here.
Because the explicit dependence on the parameter $s$ is confined in the
probability distribution $P(x_0|s)=\exp[-(x_0-\sqrt{2}s)^2]/\sqrt{\pi}$,
it is not the variables $v_1, \cdots, v_N$ but $x_0$ that is affected by
the value of the signal $s=\pm\alpha$.

By using the maximum likelihood method,
a $N$-dimensional boundary hyper-surface on the $(N+1)$-dimensional
parameter space $\{(x_0,{\bf v}^{(N)})\in {\bf R}^{N+1}\}$
that divides the likelihood regions belonging to the values
$s = \pm \alpha \in {\bf R}$
is determined by
\begin{equation}\label{eq:comp2}
P(x_0, {\bf v}^{(N)} | \alpha)p(\alpha)
=
P(x_0, {\bf v}^{(N)} | -\alpha)p(-\alpha).
\end{equation}
Here we note that the input signal $s = \pm \alpha$ appears
in the probability $P(x_0, {\bf v}^{(N)} | s)$
only through the overall factor $\exp[-(x_0-\sqrt{2}s)^2]$,
which is the same as Eq.~(\ref{eq:factorization}).
Then, the boundary condition (\ref{eq:comp2}) does not affect
the variables ${\bf v}^{(N)}$,
although does determine the threshold point of $x_0$ as
$x_0 = x_{th}$, which is the same value as that in the simplest homodyne case.

Now, let us calculate the bit error rate $BER_N'$ in the feedforward control.
The definition 
\begin{eqnarray}
BER'_N &\equiv& 
\int_{-\infty}^\infty d{\bf v}^{(N)} 
\Biggl[
       \int_{-\infty}^{x_{th}} dx_0
       P(x_0, {\bf v}^{(N)}| \alpha) p(\alpha)
\nonumber\\
&&  + \int_{x_{th}}^\infty    dx_0
       P(x_0, {\bf v}^{(N)}| -\alpha) p(-\alpha)
   \Biggr]
\end{eqnarray}
is the same as that of $BER_N$ except the coefficient $\psi_n$ of the probability
density is a function of $x_0$ and ${\bf v}^{(n-1)}$ for $n=2,\cdots,N$,
as seen from Eqs.~(\ref{eq:func-rel}).
We perform the integrations from $v_N$ to $v_1$ in this ordering,
using the normalization conditions
$\langle \psi_n | \psi_n \rangle = 1$
for $n=N, \cdots, 1$, and then integrate $x_0$
with the threshold point (\ref{eq:threshold}).

More concretely, let us consider the first integral $\int dv_N$.
From Eq.~(\ref{eq:func-rel})  the coefficients $\psi_2,\cdots,\psi_N$
does not depend on the variable $v_N$,
although it does depend on the variables $v_1,\dots,v_{N-1}$.
Then, we can carry out the integration $\int dv_{N}$
by using the normalization condition $\langle\psi_N|\psi_N\rangle = 1$,
and we eliminate the coefficient $\psi_N$.
The same step is repeated for  $v_{N-1}, \cdots, v_1$,
and then we eliminate the coefficients $\psi_{N-1}, \cdots, \psi_1$.
Finally, there remains one integral $\int dx_0$.
Now, this integration becomes the same one as that without feedforward
and that without ancillary states (the simplest homodyne case).
Finally, we obtain the same formula (\ref{eq:BERN}).


\section{Discussion}

The invariance proof relies on the transformation property of
homodyne measurement base, i.e., the eigenstates of
the quadrature amplitude and phase.
The tensor product of those bases are transformed into a separable state
through the beam splitter
\begin{eqnarray}
&& |x'\rangle_a \otimes |y'\rangle_b =
  \hat{B}_{ab}(\theta) |x\rangle_a \otimes |y\rangle_b = \nonumber \\
&&  |x\cos\theta + y\sin\theta \rangle_a
    \otimes
    |y\cos\theta - x\sin\theta \rangle_b, \label{eq:BS}
\end{eqnarray}
where $\hat{B}_{ab}(\theta) =
      \exp[i\theta(\hat{x}_a\hat{p}_b - \hat{p}_a\hat{x}_b)]$.
This is sharply contrasted to the beam splitter transformation property
of the number state bases.
For example,
\begin{equation}
\hat{B}_{ab} |1\rangle_a \otimes |0\rangle_b =
    \cos\theta |1\rangle_a \otimes |0\rangle_b +
    \sin\theta |0\rangle_a \otimes |1\rangle_b,
\end{equation}
and the output states are entangled.
By the very fact of Eq.~(\ref{eq:BS}), the homodyne distributions of
the $(N+1)$ mode systems of the signal and generic ancillae can be reduced
to the same distribution of the homodyne detection without any ancilla.
This shows that the homodyne detectors fail in causing any useful
quantum interference to reduce the BER below the homodyne limit.
Thus, the present setup does not have any necessary non-linearity
to improve the BER performance.

The future topic remained is to take into account an adaptive 
updating of the local oscillator phase at each homodyne measurement. 
Let us briefly look it at a glance. 
Varying the local oscillator phase in our model is equivalently 
realized by generalizing the beam splitting operation from 
the $SO(2)$ group to the $U(2)$ group. 
One of the most general parametrization of the $U(2)$ group elemtnt is
\begin{eqnarray}
U &=&
  e^{i\delta}e^{i\phi\sigma_z}e^{i\theta\sigma_y}e^{i\chi\sigma_z} \nonumber \\
&=&
  e^{i\delta}\left(
\begin{array}{cc}
  e^{ i(\phi+\chi)}\cos\theta & -e^{ i(\phi-\chi)}\sin\theta \\
  e^{-i(\phi-\chi)}\sin\theta &  e^{-i(\phi+\chi)}\cos\theta
\end{array}\right), \label{eq:U}
\end{eqnarray}
where $\sigma_x, \sigma_y, \sigma_z$ are the Pauli matrices.
In the two port case of the simplest example in which we have only
 one ancilla and one signal, let us adopt the most general beam splitter
$\hat{B}(U)$, the wave function becomes
\begin{eqnarray}
|\Psi' \rangle &=& \hat{B}(U) | \Psi \rangle =
\int \frac{d\alpha}{\pi}
|s'\rangle \otimes |\alpha'\rangle \psi(\alpha), \nonumber \\
\left(\begin{array}{c} s' \\ \alpha' \end{array}\right) &=&
U
\left(\begin{array}{c} s \\ \alpha \end{array}\right).
\end{eqnarray}
Next, let us consider the arbitrary phase in the each homodyne measurement
of the signal and the ancilla.
The phase rotated coordinate base of the homodyne measurement is given as
$\langle x(\phi)|\hat{q}_\phi=x\langle x(\phi)|$ where $\hat{q}_\phi =
\hat{q}\cos\phi + \hat{p}\sin\phi$.
Then, we obtain
\begin{eqnarray}
&&  P(x(\phi_0),y(\phi_1)|s) =
    |\langle x(\phi_0),y(\phi_1)|\Psi' \rangle |^2 \nonumber \\
&& =
  \int \frac{d\alpha d\beta}{\pi^2}\psi(\alpha)^*\psi(\beta)
 e^{ K - \left(x - s'' \right)^2 - \left(y - \alpha'' \right)^2 }, \label{eq:2port}
\end{eqnarray}
where
\begin{eqnarray*}
K &=& \frac{1}{2}\left( \alpha^*\beta - \alpha\beta^* \right)
- \frac{1}{2}\left| \alpha - \beta \right|^2 \\
s'' &=& \frac{ s'(\alpha)^*e^{-i\phi_0} + s'(\beta)e^{i\phi_0} }{\sqrt{2}}, \\
\alpha'' &=& \frac{\alpha'^*e^{-i\phi_1} + \beta'e^{i\phi_1}}{\sqrt{2}}.
\end{eqnarray*}
Notice that the phase dependence of a local oscillator, say $\phi$,
is translated in terms of the phase of the coherent state as
$\langle x(\phi)|\alpha \rangle = \langle x|\alpha e^{i\phi} \rangle$.
The rotation of the homodyne outcomes $x$, $y$
and the reparametrization of the phases of the homodyne measurements
forms a group $SO(2)\times U(1)^2$.
If we could reparametrize $x,y,\phi_0, \phi_1$ to eliminate the beam splitter
parameter $U$ in the two port probability (\ref{eq:2port}),
the methodology used in this letter would be directly applicable and 
thus would obtained the same conclusion; namely, the BER is invariant.
However, it is not the case in actual, that is, we cannot absorb all of 
these parameters in Eq.~(\ref{eq:U}).
Precisely, the unabsorbed parameters is in the coset group
$U(2)/(SO(2)\times U(1)^2)$. 
The remained question is to find a way of treating them 
to clarify whether this parameter space helps us 
to reduce the bit error rate further
than that of a simple homodyne detection.


\newcommand{\Nature} [3]{{\it Nature}           {\bf #1}, #2 (#3)}
\newcommand{\NatPhys}[3]{{\it Nature Phys.}     {\bf #1}, #2 (#3)}
\newcommand{\PR}     [3]{{\it Phys. Rev.}       {\bf #1}, #2 (#3)}
\newcommand{\PRL}    [3]{{\it Phys. Rev. Lett.} {\bf #1}, #2 (#3)}
\newcommand{\OL}     [3]{{\it Opt. Lett.}       {\bf #1}, #2 (#3)}
\newcommand{\OEx}    [3]{{\it Opt. Ex.}         {\bf #1}, #2 (#3)}
\newcommand{\Science}[3]{{\it Science}          {\bf #1}, #2 (#3)}
\newcommand{\RMP}    [3]{{\it Rev. Mod. Phys.}  {\bf #1}, #2 (#3)}

\end{document}